\documentstyle[aps,prl,epsfig,multicol,color]{revtex} 
\newenvironment{Figure}{\vskip 0pt\noindent}{\vskip 0pt\noindent} 
\newcommand{\Caption}[1]{\par\noindent{\small #1} \par} 
 
\newcommand{\Beginrule}{\vskip 3pt\noindent\hbox{%
\vbox{\hbox to 9cm{\hfill}}\vbox{\hrule width 9cm}} \vskip 3pt}
\newcommand{\infig}[2]{\begin{center}{\epsfig{file=#2,width=#1}}\end{center}} 
\begin{document}

\title{Role of quasiparticles in the growth of a trapped Bose-Einstein 
condensate}
\author{C.W.~Gardiner$^{1}$ M.D. Lee$^{1}$  R.J. Ballagh$ ^{2}$ M.J. Davis$ 
^{2}$ 
and P. Zoller$ ^{3}$.}
\address{$^1$ School of Chemical and Physical Sciences, 
Victoria University, Wellington, New Zealand}
\address{$^2$ Physics Department, University of Otago, Dunedin, New 
Zealand}
\address{$^3$ Institut f\"ur theoretische Physik, Universit\"at 
Innsbruck, A6020 Innsbruck, Austria.} \maketitle

\begin{abstract}
A major extension of the model of condensate growth introduced by us is 
made to take account of the evolution of the occupations of lower trap 
levels (quasiparticles) by scattering processes, and of the full 
Bose-Einstein formula for the occupations of higher trap levels, which 
are assumed to have a time independent occupation.  The principal 
effect is a speedup of the growth rate by somewhat less than an order 
of magnitude, the precise value depending on the the assumptions made 
on scattering and transition rates for the quasiparticle levels.
\end{abstract}

\pacs{PACS Nos. }
\long\def\comment#1{\vskip 2mm\noindent\fbox{%
\vbox{\parindent=0cm{\small\em #1}}}\vskip 2mm}
\begin{multicols}{2}

In a previous paper \cite{BosGro}, we introduced a formula for the 
growth of a Bose-Einstein condensate, in which growth resulted 
exclusively from stimulated collisions of atoms where one of the atoms 
is left in the condensate.  This gave a very simple growth formula 
which predicts a rate of growth of the order of magnitude of that 
observed in current experiments \cite{JILA,MIT,RICE}. The direct 
stimulated effect must eventually be very important, once a 
significant amount of condensate has formed, but in the initial stages 
there will of course also be a significant number of transitions to 
excited states of the condensate (quasiparticles), whose populations 
will then also grow.  As well as this, there will be interactions 
between the condensate, the quasiparticles and the atomic vapor from 
which the condensate forms.  This paper extends the description of the 
condensate growth to include these factors.

All of these effects are encompassed by the description given in 
\cite{BosGro,QKIII}.  However, the practical extension of this 
description to take account of the additional effects would involve a 
calculation of all the eigenfunctions for the trapped condensate, and the 
detailed summation over all processes involving these.  Since 
the number of levels involved is of the order of tens of thousands, 
this could be a formidable task.  However some quite reasonable estimates 
can be made for the overall effects of these processes, and from 
these we can derive a set of easily solvable differential equations 
for the populations of the condensate and the lower energy 
quasiparticles.

As in our previous work, we divide the states in the potential into 
the {\em condensate band}, $ R_C$, which consists of the energy levels 
significantly affected by the presence of a condensate in the ground 
state, and the {\em non-condensate band}, $ R_{NC}$, which contains all the 
remaining energy levels above the condensate band.  The division 
between the two bands is taken to be at the value, $ E_R$.

The picture we shall use assumes that $ R_{NC}$ consists of a large 
``bath'' of atomic vapor, whose distribution function is given, for 
the energy levels greater than a value which we shall call $ E_{\rm 
max}$ (with $E_{\rm max} > E_{R}$), by a time-independent equilibrium 
Bose-Einstein distribution $\{\exp[(E-\mu)/k_BT] -1\}^{-1}$.  The value 
of $ E_{\rm max}$ will be assumed to be small enough for the majority 
of atoms to have energies higher than $ E_{\rm max}$, so that this part 
of the bath can be treated as being essentially undepleted by the 
process of condensate growth.

\begin{Figure}
\infig{7cm}{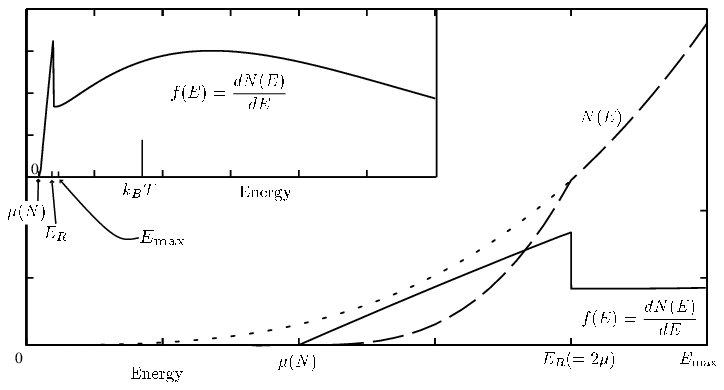} \Caption{Fig.1: Modification of the equilibrium level 
occupation by condensate growth, for a 3D harmonic trap.  
The cumulative occupation of states $N(E)$ with energy below $E$, is shown 
for a gas of noninteracting atoms as a dotted line, and after 
condensate growth (when the ground level energy has increased by 
$\mu(N)$) is shown as a dashed line.  The solid line gives the 
occupation $f(E) = \frac{dN(E)}{dE}$ after condensate growth.  The 
values of $\mu,\mu(N)$, and $k_BT$ correspond to the MIT sodium 
condensate of $10^6$ atoms at 1.2 $\mu K$.
}
\end{Figure}

The energy levels between $ E_R$ and $ E_{\rm max}$ are taken to have a 
time dependent population, since the continuation of the equilibrium 
Bose-Einstein formula to lower energies eventually leads to an unrealistic 
situation in which the populations and transition rates become too large 
for the populations to be considered to be constant---this demonstrates 
that the initial condition in which the vapor is at a positive chemical 
potential cannot apply for all energies.  The choice of $ E_{\rm max}$ is 
thus determined as a lower limit to the equilibrium distribution, with the 
distribution in the range between $ E_{\rm max}$ and $ E_{R}$ treated 
as time dependent, and computed as part of the growth kinetics.

The value $ E_R$, above which the energy levels are taken as 
unperturbed, was fixed at $ 2\mu$.  This value and the ground state 
energy level---the chemical potential $ \mu(N)$---put bounds on the 
energy levels of the states in between.  As a simple expression of 
this fact, the levels between $ \mu(N)$ and $ E_R$ are determined by 
interpolating linearly between the two extremes, using a density of 
states ${\cal N}[E-\mu(N)]^{2}$, where ${\cal N}$ is a normalization 
chosen so that the cumulative number of states matches the 
corresponding cumulative value for an unperturbed three dimensional 
harmonic oscillator when $E=E_{R}$.  As an illustration of the effect 
of this, we show in Fig.1 the cumulative occupation $N(E)$, and the 
occupation per unit energy interval $f(E)=dN(E)/dE$, when the system 
is in equilibrium.  (The condensate population itself is not shown.)

\begin{Figure}
\infig{7cm}{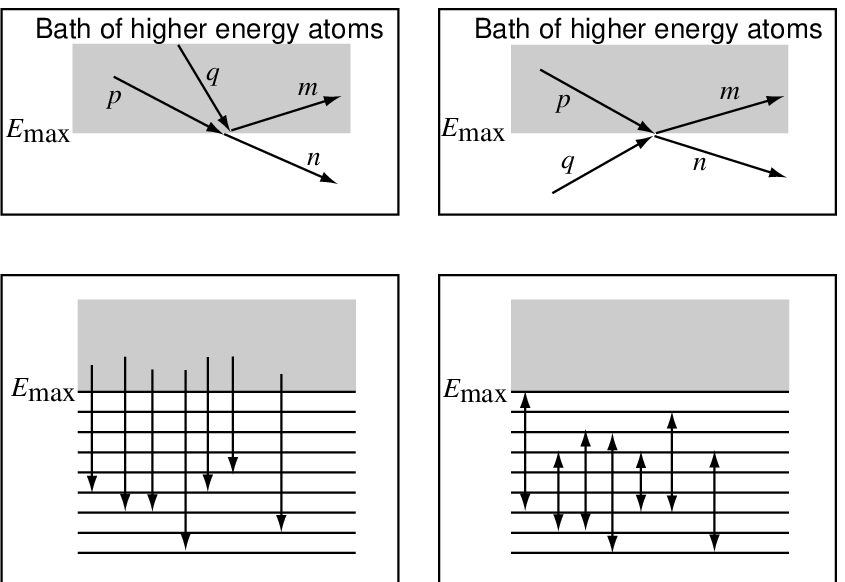} \Caption{Fig.2: The transitions being 
considered: Left---scattering; Right---Condensate growth.}
\end{Figure}

The dynamics we will consider will arise from two 
kinds of process as illustrated in Fig.2.

\noindent {\em Growth}: A collision between a pair of atoms initially 
in the bath of atomic vapor results in {\em one} of the atoms having a 
final energy less than $ E_{\rm max}$.

\noindent
{\em Scattering}: A collision between an atom initially in an energy 
level below $ E_{\rm max}$ and a bath atom transfers the first atom 
to another energy level below  $ E_{\rm max}$.

Our treatment therefore omits any scattering between atoms which both 
have energies less than $ E_{\rm max}$, which is reasonable if the 
number of atoms in the bath is almost 100\% of the total number 
of atoms.

The mechanism of condensate growth, as in \cite{BosGro,QKIII}, under 
appropriate approximations, gives rise to equations of motion for the 
number of atoms in the condensate band $ N$, and the number of 
quasiparticle excitations $ n_m$ (with energies $ e_m<E_{\rm max}$) in the 
condensate, which can be written as follows.  First define
\begin{eqnarray}\label{phys1}
\dot n^+_m 
&\equiv& 
2W^{++}_m(N)\left[(1-e^{(\mu(N)-\mu+e_m)/k_{B}T})n_m+1\right],
\\ \label{phys2}
\dot n^-_m 
&\equiv& 
2W^{-+}_m(N)\left[(1-e^{(-\mu(N)+\mu+e_m)/k_{B}T})n_m+1\right].
\end{eqnarray}
Then the multilevel growth equations are
\begin{eqnarray}\label{phys3}
\dot n_m &=& \dot n^+_m +\dot n^-_m,
\\ \dot N &=& 2W^+(N)\left((1-e^{(\mu(N)-\mu)/k_{B}T})N+1\right)
\nonumber  \\&&
+\sum_m\left\{ \dot n^+_m -\dot n^-_m\right\}.
\end{eqnarray}
Here $2(N+1)W^+(N)$ is the transition rate for an atom to enter the 
condensate level as a result of a collision between two atoms in the 
``bath'' of atomic vapor---the factor $N+1$, takes account of both the 
``spontaneous'' term, and the ``stimulated'' term induced by the 
presence of the condensate.  The reversed process occurs with the 
overall rate $2NW^+(N)e^{(\mu(N)-\mu)/k_{B}T}$---that is with no 
``spontaneous'' term, and with a factor dependent on the difference of 
the chemical potential $\mu$ of the vapor, and that $\mu(N)$ of the 
condensate.  As a result, equilibrium in the large $N$ limit occurs at 
equality of the two chemical potentials.

During the process of BEC formation, the spectrum of eigenvalues makes 
a transition from the unperturbed spectrum of trap levels to the case 
where the spectrum is strongly affected by the condensate in the 
ground state.  The Bogoliubov spectrum of a condensed gas is valid in 
the case where the number of particles in the condensate, $ n_0$, is 
so large that it is valid to write $ n_0 \approx N$.  Thus, during the 
initial stages of condensate formation, where this is not true, one 
must use another formalism.  In this paper we will consider the 
situation in which the interaction between the particles is very weak, 
as is in practice the case.  This means that we will be able to use 
the unperturbed spectrum for the initial stages of condensation, and 
only use the Bogoliubov description once enough condensate has formed 
to make the effective interaction rather stronger.

The basic formalism of \cite{QKIII} can still be carried out in this 
case, and the modification that is found is rather 
minor---essentially, we make the substitution $ N\to n_0$ in the 
chemical potential and the $ W^+(N), W^{++}(N)$ functions, and set 
$W^{-+}_m\to 0 $, since this term comes from the mixing of creation 
and annihilation operators which arises from the Bogoliubov method.  
In order to simplify the equations we also group the levels in narrow 
bands of energy with $ g_k$ levels per group, and for simplicity use 
the same notation $ n_k$ now for the number of particles in the energy 
band with mean energy $ e_k$.  (This corresponds to the {\em ergodic} 
assumption used in \cite{Holland KE}.)  We then deduce
\begin{eqnarray}\label{phys4}
\dot n_m|_{\rm growth} 
&\equiv& 
2W^{++}_m(n_0)\left\{\left[1-e^{\mu(n_0)-\mu+ e_m\over k_{B}T}\right]n_m+g_m
\right\},
\nonumber \\
\\ \label{phys5}
\dot n_0|_{\rm growth}  &=& 
2W^+(n_0)\left\{\left[1-e^{\mu(n_0)-\mu\over k_{B}T}\right]n_0+1
\right\}.
\end{eqnarray}
The growth equations can be modified to include the terms derived in 
\cite{QKIII} corresponding to the scattering of particles in the 
condensate band by the vapor particles.  This is equivalent to 
scattering of particles by a heat bath, which leads to a rate equation 
for scattering of the form (Where $\bar N_{km}= 1/(\exp[(e_k -e_m)/k_BT]-1)$, 
and by $ k>m$ we mean $ e_k > e_m$)
\begin{eqnarray}\label{growth5}
&& \dot n_m|_{\rm scatt} =
\nonumber\\
&& \sum_{k<m}\!\!
\gamma_{mk}
\left\{\bar N_{mk}n_k(n_m+g_m) -(\bar N_{mk}+1)(n_k+g_k)n_m\right\}
\nonumber \\ 
&& + \!\!\sum_{k>m}\!\!
\gamma_{km}
\left\{(\bar N_{km}+1)n_k(n_m+g_m) -\bar N_{km}(n_k+g_k)n_m\right\}.
\nonumber \\
\end{eqnarray}
The formulae of QKIII give precise methods for computing the 
coefficients $ \gamma_{km}$, but we can simplify their computation by 
adapting the kinetic equation of Holland {\em et al.} 
\cite{Holland KE}.  This methodology is based on a model in which the 
trap levels are all treated as being unaffected by the presence of the 
condensate, which should suffice as a first approximation.  To apply 
it to this situation, we assume all the levels with energies greater 
than $ E_{\rm max}$ are thermalized,
and sum out over these levels.  The working is essentially 
straightforward, and yields an equation for the $ n_m$ variables in 
the form (with $M$ the mass of the atom and $a$ the scattering length)
\begin{eqnarray}\label{growth14}
&&\left.{\dot n_m }\right |_{\rm scatt}
={8Ma^2\omega^2\over\pi\hbar } e^{\mu/k_BT}\Gamma(T)\times
\nonumber\\
&&\,\,
\Bigg\{\sum_{k<m}{1\over g_m}\left[n_k(g_m+n_m)
{e^{-\hbar\omega_{mk}/k_BT} }
-n_m(g_k+n_k)\right]
\nonumber \\
 &&\,\,  +
\sum_{k>m}{1\over g_k }\left[n_k(g_m+n_m)
-n_m(g_k+n_k){e^{-\hbar\omega_{km}/k_BT}}\right]\Bigg\}.
\nonumber \\
\end{eqnarray}
where $ \Gamma(T) \equiv\sum_{e_m>e_{\rm max}}e^{-e_m/k_BT} $
 has a value which depends on the spectrum of 
energies.  For an {\em isotropic} 3-dimensional harmonic oscillator with 
frequency $ \omega$, the energy levels above the zero point are 
$ e_n = n\hbar\omega$, so that we find
\begin{eqnarray}\label{growth10a}
\Gamma(T) ={ e^{-E_{\rm max}/k_BT}\over1-e^{-\hbar\omega/k_BT}}.
\end{eqnarray}

This corresponds to essentially to (\ref{growth5}) when one makes the 
correspondences
\begin{eqnarray}\label{growth15}
&&\bar N_{km} \to e^{-(e_{k}-e_{m})/k_BT},\quad 1 +\bar N_{km} \to 1 \\
&&\gamma_{km}=\gamma_{mk}\to {8Ma^2\omega^2\over\pi\hbar } { 
e^{\mu/k_BT}\Gamma(T)\over g_k} \quad\mbox{with } k>m.
\end{eqnarray}

The equation for both growth and scattering is now given by
adding (\ref{phys4}) to (\ref{growth15})
\begin{eqnarray}\label{growth1501}
\dot n_{m} &=& \dot n_{m}|_{\rm growth} + \dot n_{m}|_{\rm scatt},
\end{eqnarray}
where, for $ m=0$, we use (\ref{phys5}) instead of (\ref{phys4}).

The overall evolution of the system can now be found from the 
numerical solutions to (\ref{growth1501}).  The parameters used were 
chosen so as to be in approximate agreement with the experimental 
work being conducted at MIT, where the growth of Bose-Einstein condensates of 
$^{23}{\rm Na}$ is being studied.  In contrast to the estimate in 
\cite{BosGro} in which the bath distribution was approximated by a 
Maxwell-Boltzmann distribution, in this computation we use the full 
Bose-Einstein distribution, truncated at a lower energy of $ E_{\rm max}$, 
since lower energies are described by the $ n_m$ variables.

In applying the theory two major approximations 
are made.  Firstly, the $W_m^{++}(N)$ functions were assumed to 
be equal to the $W^+(N)$ function, since the actual $W_m^{++}(N)$ 
terms are not easily calculated.  The justification for this is that 
the $W_m^{++}(N)$ terms represent an averaging over all 
the levels contained in the $m$th group, and as such they may be 
expected to be of the same order of magnitude as $W^+(N)$.  As a 
validity check, it was found that the effect on the 
condensate growth rate was small when the $W_m^{++}(N)$ terms were 
altered by a factor in the range $0.5 - 2$.

In current BEC experiments the confining harmonic potential is 
normally anisotropic, whereas \cite{Holland KE} was restricted 
to an isotropic trap.  The second approximation is therefore that the 
scattering rate factor $\Gamma(T)$ is equal to that for an isotropic 
3D harmonic oscillator with frequency equal to the geometric mean 
frequency of the anisotropic trap.  The precise value of this factor 
was found to have little effect on the solutions, so long as it was 
greater than about one tenth of the value given by (\ref{growth10a}).

The condensate rate of growth depends on the number of groups of 
levels considered in the model---modeling more groups in the 
condensate band increases the rate of growth, which approaches a 
limiting value.  The number of groups modeled was therefore chosen as 
large as possible, but it was required that there were at least 4 
levels in the first group of levels above the condensate level.
\begin{Figure} 
\infig{8.4 cm}{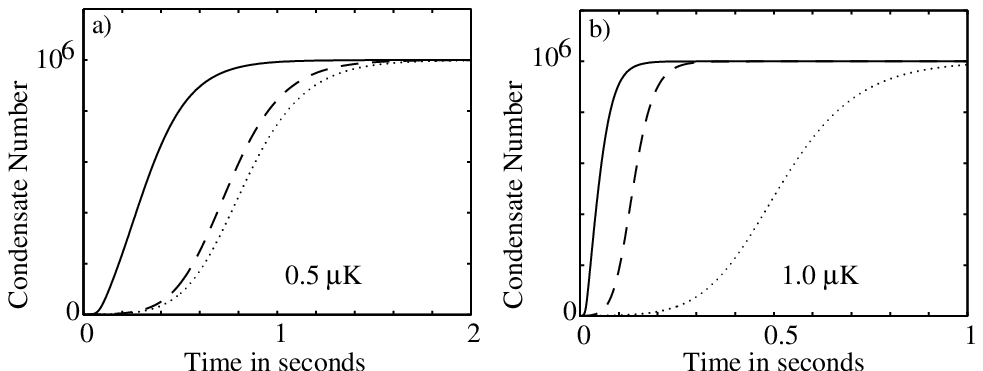}
\Caption{Fig.3: Condensate growth for Sodium: Dotted line---Growth of the 
condensate for the uncorrected 
model of \cite{BosGro}; Dashed line---scattering is neglected, but  
$ W^+$ is given by using the Bose-Einstein correction; Solid line---with 
scattering and the Bose-Einstein correction.  In all cases the initial amount 
of condensate at $ t=0$ was 500 atoms. }
\end{Figure}%
\noindent%
The initial populations for the groups in the condensate band were 
generated by a method which models the experimental procedure.  We 
start at some initial $t\ll 0$ with the bath at a chemical potential 
$\mu \approx 0$, and evolve the equations of motion until equilibrium 
is reached; at $t=0$ we then change $\mu$ to a positive value such 
that $\mu = \mu(N_{\rm final})$, where $N_{\rm final}$ is the final 
number of atoms in the condensate.  Changing the populations at $t=0$ 
merely moves the growth curve slightly forward or backward in time, 
with no effect on its basic shape.  We present a sample of the results 
obtained in Fig.3.  The trap parameters, $ \nu_x=18.5{\rm 
Hz},\nu_y=\nu_z=250{\rm Hz}$, temperatures and final condensate number 
are chosen in the range presently being investigated for sodium.

The typical behavior of the noncondensate levels is shown in Fig.4, 
for the case where the initial occupation of all groups below $ E_R$ 
is chosen to be zero, in order to show the speed of the relaxation 
process.  The initial growth is in the population of the noncondensate 
levels---their occupations can become very large, but this is 
because the numbers of levels in each group are very large, so that 
there is little degeneracy, i.e., the number of atoms per {\em level} 
does not significantly exceed one.  Even in the lowest group the 
degeneracy is no more that about 100.  The moment the condensate 
achieves a significant degeneracy, the stimulated process takes over, 
and immediately draws the excess population of the 
noncondensate groups into the condensate, well before the full 
condensate occupation is achieved.  Thus, apart from the initial 
transient, the condensate growth occurs essentially by the same 
mechanism as in \cite{BosGro}, with the modification that the 
distribution over the noncondensate levels changes slowly in response 
to the change of the condensate chemical potential.

If scattering is entirely neglected the populations of the lowest 
noncondensate levels become several times larger than those of the the 
condensate before settling to their very much lower equilibrium 
values.  However, the inclusion of even as little as 0.1\%
of the strength of scattering used here eliminates that effect almost 
entirely.

\begin{Figure}
\infig{4.5cm}{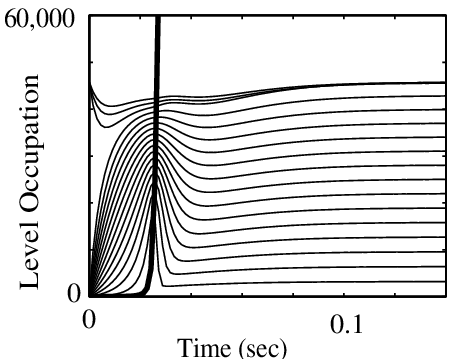}
\Caption{Fig.4: Growth of noncondensate levels---the condensate number is the 
almost vertical black line.}
\end{Figure}

\noindent%
In contrast to our work, in which explicit use is made of trap 
eigenfunctions, other descriptions \cite{Older,Kagan97,Stoof97} of 
condensate growth either do not treat the case of a trapping 
potential, or consider only the case where trapped situation is not 
essentially different from the untrapped situation.  Nevertheless, the 
equations we use have a strong connection with those of Stoof 
\cite{Stoof97}, but their interpretation is different, since 
Stoof's Fokker-Planck equation treats the untrapped case, where the 
low-lying levels are extremely closely spaced, and relative coherences 
between these levels become important.  The current experiments are 
probably closer to the situation of our model, since the growth rate 
is somewhat slower than the lowest trap frequency.

We can conclude from the results of this model that the main effect of the 
inclusion of the scattering and the computation of $  W^+$ using the 
full Bose-Einstein formula is to speed up the condensate growth by up to one 
order of magnitude, the precise speedup depending on the condensate size and 
temperature.  Even though we have only estimated the scattering and 
transition rates for the quasiparticle levels, the results are reasonably 
predictive, since it would be hard to credit the scattering or the $ W^{++}$ 
factors as being very different from the values assumed, and also because the 
numerical predictions are not very sensitive to these precise values.  Precise 
predictions will involve the detailed computation of these effects, rather 
than their estimation.

\noindent{\bf Acknowledgments:}
We would like to thank Wolfgang 
Ketterle and Hans-Joachim Miesner for discussions regarding sodium 
experiments, and Eric Cornell, Carl Wieman, Deborah Jin and Jason 
Ensher for discussions regarding rubidium experiments.  This work was 
supported by the Marsden Fund under contract number PVT-603, and by 
{\"O}sterreichische Fonds zur F{\"o}rderung der wissenschaftlichen 
Forschung.

\end{multicols}


\begin{references}
\bibitem{BosGro} C.W.~Gardiner, P.~Zoller, R.J.~Ballagh and 
M.J.~Davis, Phys. Rev. Lett. {\bf 79}, 1793 (1997).

\bibitem{JILA} M. Anderson, J.R. Ensher, M.R. Matthews,
C.E. Wieman and E.A. Cornell,
Science {\bf 269}, 198 (1995).

\bibitem{MIT} K.B. Davis, M-O.Mewes. M.R. Andrews. N.J. van Druten, D.S.
Durfee, D.M. Kurn, and W. Ketterle,
Phys. Rev. Lett. {\bf 75}, 3969 (1995).

\bibitem{RICE} C.C. Bradley, C.A. Sackett, J.J. Tollet, and  R. Hulet,
Phys. Rev. Lett. {\bf 75}, 1687 (1995).

\bibitem{QKIII}{C.W. Gardiner} and {P. Zoller}, cond-mat/9712002.

\bibitem{Holland KE} M. Holland, J. Williams, K. 
Oakley, J. Cooper, Phys. Rev. A {\bf 55}, 3670 (1997)

\bibitem{Older} E. Levich and V. Yakhot, Phys. Rev. B {\bf 15}, 243 
(1977); 
J.Phys.A {\bf 11}, 2237 (1978);
%
D.W. Snoke and J.P. Wolfe, Phys. Rev. B {\bf 39}, 4030 (1989);
%
H.T.C. Stoof, Phys. Rev. Lett. {\bf 66}, 3148 (1991); 
%
Yu. M. Kagan, B. V. Svistunov and G.V. Shlyapnikov,
Sov. Phys JETP {\bf 75}, 387 (1992);
%
D.V.  Semikoz and I.I.  Tkachev, Phys.  Rev.  Lett.  {\bf 74}, 3093 
(1995); %
H.T.C. Stoof, Phys. Rev. A {\bf 49}, 3824 (1994).
\bibitem{Kagan97} Yu. Kagan and B. V. Svistunov, Phys. Rev. {\bf 79}, 3331 
(1997).
\bibitem{Stoof97} H.T.C. Stoof, Phys. Rev. Lett. {\bf 78}, 768 (1997).
\end{references}
\end{document}